\documentclass{article}

\newcommand{\gtsimeq}{\raisebox{-0.6ex}{$\,\stackrel
{\raisebox{-.2ex}{$\textstyle >$}}{\sim}\,$}}

\begin{document}

\begin{center}
\large{\textbf{A Population of Very Young Brown Dwarfs and Free-floating 
Planets in Orion}}
\end{center}

\normalsize


\begin{center}
\textbf{P.W.Lucas$^{1,2}$ and P.F.Roche$^2$}
\end{center}

Accepted by Monthly Notices of the Royal Astronomical Society



\begin{center}
\textbf{Abstract}
\end{center}

        We describe the results of a very deep imaging survey of the Trapezium
Cluster in the $IJH$ bands, using the UKIRT high resolution camera UFTI.
Approximately 32\% of the 515 point sources detected are brown dwarf candidates, 
including several free floating objects with masses below the Deuterium burning 
(planetary) threshold at 0.013 solar masses, which are detectable because of 
their extreme youth. We have confidence that almost all the sources detected are 
cluster members, since foreground contamination is minimal in the 33 arcmin$^{2}$ area 
surveyed and the dense backdrop of OMC-1 obscures all background stars at these 
wavelengths. Extinction is calculated from the \it{(J-H)} \rm colours, permitting 
accurate luminosity estimates and temperatures are derived from the dereddened 
\it{(I-J)} \rm colours. There is some evidence for a cut-off in the luminosity 
function below the level corresponding to several Jupiter masses, which may 
represent the bottom end of the IMF. Since star formation is complete in the 
Trapezium this limit could have wide significance, if confirmed. However, it 
could well be an effect of the dispersal of the molecular cloud by the central 
O-type stars, a process whose timescale will vary between star formation regions. 

\vspace{3mm}
(1) Dept. of Physical Sciences, University of Hertfordshire, College Lane,
Hatfield AL10 9AB.\\ 
(2) Astrophysics Dept., University of Oxford, 1 Keble Road, Oxford OX1 3RH.





\section{Introduction}

	The last five years have seen great progress in the detection 
of brown dwarfs in the Local Neighborhood, young Galactic Clusters
and star formation regions, starting with the near simultaneous discovery of
the first clearly confirmed brown dwarfs, Teide 1 in the Pleiades (Rebolo,
Zapatero-Osorio \& Martin 1995) and Gl229b in the Local Neighborhood 
(Nakajima et al.1995). Star formation regions offer the advantage 
that substellar objects are 3 orders of magnitude more luminous at an age of
a few Myr than at an age of a few Gyr. Early photometric and spectroscopic work 
(Comeron et al.1993,1996; Williams et al.1995) indicated that brown dwarfs are 
probably very common in star formation regions. However, confirmation of substellar
status is problematic in star formation regions, owing to the ubiquity of
Lithium in young objects and the complicating effects of extinction on both
photometry and spectroscopy. 

	Recently, high quality spectroscopy (Luhman \& Rieke 1998; Luhman et al. 1998,
Wilking, Greene \& Meyer 1999) and the publication of theoretical evolutionary models 
for young substellar objects (Burrows 1997;D'Antona \& Mazzitelli 1998, hereafter 
B97 and DM98) have provided convincing evidence that photometric identification of 
young brown dwarf candidates is reliable. In photometric studies, masses of candidate 
objects are derived by comparison of the observables (luminosity and temperature) with 
the evolutionary tracks. The isochrones of B97 and DM98 are in fairly good agreement in
regard to the mass-luminosity relation at an age of about 1 Myr but there is 
some disagreement about the mass-temperature relation (HR diagrams are compared
by Luhman \& Rieke 1998). Even if the theoretical effective temperatures
were without flaw, there is considerable uncertainty in the derivation of temperatures
from photometry or spectroscopy, at the level of $\pm 200~K$ in M and L dwarfs.
Hence, we use luminosity, which is more easily measured, to derive masses for
our sources.

	In this paper we report the results of a deep infrared photometric survey
of the Trapezium Cluster in Orion. A large population of substellar objects is
discovered, including the first free-floating objects of planetary mass. We
note that the IAC group (Bejar et al.1999, not yet published) has simultaneously 
reported a similar discovery of planetary mass objects in the adjacent $\sigma$ 
Orionis 
cluster. The Trapezium has been intensively studied for many years and we have been
able to draw upon a large body of publications to aid in our work. We selected
the Trapezium for several reasons. (1) Its very high stellar density allowed
photometry of several hundred sources in a fairly small survey. (2) False positive 
detections are essentially eliminated because the dense backdrop of OMC-1 obscures all 
background stars even at K band, as shown by Hillenbrand \& Hartmann (1998) through 
optical-infrared comparison of the cluster stellar density profile. (3) The extinction 
within the cluster is relatively low ($0 < A_{V} < 15$ for most sources), permitting
reasonably precise dereddening. (4) Star formation is essentially complete in the 
cluster and the age range is thought to be 0.3-2~Myr, so the age-luminosity degeneracy
is not large.


\section{Observations}

	Deep Imaging of the Trapezium cluster was carried out at United Kingdom 
Infrared 
Telescope (UKIRT) on 14-16 December and 22-23 December 1998, using UFTI, the UKIRT Fast 
Track Imager. The observations of 22-23 Dec were made by observatory staff to
compensate for time lost to poor weather and equipment failures on 15-16 Dec.
UFTI is a high resolution camera constructed by the authors at Oxford University
with assistance from several other UK institutions (Roche \& Lucas 1998).  
It has a 1024$\times$1024 HAWAII array sensitive between 0.78 and 2.5 microns.
The image scale is 0.091 arcsec/pixel, yielding a field of view of 92.6 arcsec.
Observations of 15 contiguous fields were made in the $I, J$ and $H$ filters, with 
900s exposures in each filter. Twilight flatfields were taken to reduce the data. Seeing 
conditions were typically 0.6 arcsec FWHM in all 3 filters, with only slightly poorer 
image quality at $I$ band. The fine pixel scale led to a high degree of over-sampling, 
which was very useful for distinguishing stars from small knots of nebulosity 
and in permitting reliable photometry of low signal to noise detections.
  
\subsection{Filter Selection}

	Taking advantage of the short wavelength sensitivity
of the HAWAII array UFTI contains $I_{U}$ (0.786-0.929~$\mu$m) and 
$Z_{U}$ (0.85-1.05~$\mu$m) band filters. These filters sample the steeply
rising part of brown dwarf spectra from the optical to the near infrared flux peak. 
The $I_{U}$ filter was selected because the $Z_{U}$ band contains a very strong 
[SIII] emission line which might have contaminated the photospheric flux
in ``proplyd'' sources and because the \it{(I-J)} \rm colour is a more reliable
temperature indicator than than the smaller \it{(Z-J)} \rm colour given that fluxes 
have to be dereddened. The $J$ and $H$ filters were chosen to measure extinction 
using the temperature insensitive \it{(J-H)} \rm colour. The $K$ filter was not 
selected because a significant fraction of the flux comes from hot circumstellar dust 
at 2.2~$\mu$m, making it unreliable for determining extinction or luminosity and 
because it contains fairly strong low-excitation emission lines.

	The UFTI $I_{U}$ band calibration (equ.1) relative to Cousins $I$ was made by 
observation of 8 faint red standards of approximately solar metallicity taken from Leggett 
et al.(1998). The calibration has been confirmed by observatory staff, including observations 
of blue standards, and can be fit remarkably well by a linear equation. The $I_{U}$ 
bandpass is somewhat redder than $I_{C}$, which compensates for the low quantum
 efficiency 
of the HAWAII array at these wavelengths ($\sim 23\%$) when observing cool stars.
The $J$ and $H$ filters are of the new type commissioned by the Mauna Kea Consortium.
The transformation to the CIT colour system (equ.2,3) was determined by combining
the transformations of S.Leggett for CIT to UKIRT($_{IRCAM}$) with that for 
UKIRT($_{IRCAM}$) to UKIRT($_{UFTI}$). We use the UFTI $J$ and $H$ magnitudes in this 
paper but use Cousins $I$ for ease of comparison with other studies. This is possible 
because fortuitously the net effect of extinction on equ.1 is much less than the 
measurement errors. 

\begin{equation}
I_{C} = I_{U} + 0.273(I_{U} - J_{U}) 
\end{equation}
\vspace{-4mm}
\begin{eqnarray}
(J-H)_{U} = 1.03 (J-H)_{CIT} \\
J_{U} = K_{CIT} + 1.132(J_{CIT} - K_{CIT})
\end{eqnarray}
\vspace{-1cm}
\subsection{Data Reduction and Photometry}

	The data were reduced using IRAF and the Starlink package CCDPACK
for image mosaicing. Photometry was carried out using the DAOPHOT package in IRAF.
Photometry in the core of the Trapezium cluster is complicated by the pervasive
nebulosity, which has structure on all the observed spatial scales. This often
leads to inaccurate measurement of sky background when doing automated photometry,
and causes the DAOFIND algorithm to misidentify many small-scale nebular flux
variations as stars. To overcome these problems, we cross correlated the stars found in 
each filter to remove most of the nebulous sources and also rejected all very blue 
sources ($(J-H) < 0.2$), which inspection indicated were all spurious. Every source
was then visually inspected and photometry was performed manually in order to 
select the best approporiate sky annulus in each case. The results of manual photometry 
were generally used in preference to the results of automated crowded-field photometry, 
with a 
few exceptions where the ALLSTAR routine was needed for photometry of close binaries. 
The final photometric precision is $\sim 5\%$ in the outlying regions of the survey
where the nebulosity is faint, limited by temporal variations in the image profile. 
Precision in the bright nebulous core (in which the majority of sources lie) is between 
5\% and 20\% (for the worst cases), depending on the degree of spatial variation of surface 
brightness and the source magnitude. Since a large fraction of young stars are 
variable, more precise photometry at one epoch would not have had much greater value.

\section{Results}

	The main results of the survey are presented graphically in Figure 1(a-b). 
515 unsaturated point sources were detected in both the $J$ and $H$ bands, of which 313 were also 
detected at I band. An additional 48 sources were detected at greater than 5-$\sigma$ in the $H$ 
band alone, and appear as upper limits in Figure 1(b). An approximate colour-magnitude sequence 
for zero extinction is indicated by the dotted line. The value of the nearly temperature 
independent intrinsic \it{(J-H)} \rm colour over the range $\sim$2200-4000~K is clear, 
since nearly all low mass stars and substellar sources will deredden to a colour near 
\it{(J-H)} \rm= 0.6. However, we prefer to use a two-colour sequence for formal dereddening 
where possible (see Section 3.2). The empty region to the upper right of the diagram 
represents the saturation limit near m$_{H}=11.7$.

	A large fraction ($\approx 32\%$) of the $JH$ sources are brown dwarf candidates, 
lying below the reddening track for a 0.08~M$_{\odot}$ star at an age of 1~Myr, as 
calculated from the B97 isochrones. Approximately 13 sources appear to lie below the 1~Myr 
track for an object with the minimum mass to burn Deuterium ($\approx$ 0.013~M$_{\odot}$). 
Following the definition suggested by Burrows (1997) it is convenient to call such objects 
free-floating planets, even though they are likely to have formed by cloud core 
fragmention in the same manner as stars and brown dwarfs. A definition by mass has the
advantage that it can be applied before all the formation mechanisms are known.
An interesting feature of Figure 1(a) is the paucity of very faint blue sources
with m$_{J} > 19$, given that several fairly red sources are seen below this limit. This 
observation is based on a small number of objects but it is supported by the upper limits in
Figure 1(b), which show that many very faint red sources are detected at H band only, but
none with \it{(J-H)} $<1.3$.\rm This appears to indicate a sharp drop in the cluster 
Luminosity Function, at a level corresponding to about 8 M$_{Jup}$. The significance of 
this is discussed in Section 4.2.

	$IJH$ photometry of the observed point sources is presented in Table 1\footnote{
Table 1 is available electronically by FTP to star.herts.ac.uk, in pub/Lucas/Orion.} 
together with astrometry, dereddened magnitudes, luminosities, derived masses (using 
both the B97 and DM98 isochrones) and temperatures. We have surveyed the inner regions of 
the Orion Nebula cluster, in which most star formation is thought to have occurred over the
 period between 
0.3 and 2~Myrs ago (Ali \& Depoy 1995, Hillenbrand 1997), which leads to a small 
uncertainty in the derived masses, which we have indicated on Figure 1(a) by plotting the 
brown dwarf threshold for three different ages. A small proportion of younger 
sources is undoubtedly present, whose masses will be less than we have estimated, but we have
excluded the 18 sources which exhibited extended (non-stellar) profiles from our photometry 
and performed a further colour selection against proplyds (see Section 3.2) so only a few 
extremely young sources should remain in Table 1.

	The $I$ vs.\it{(I-J)} \rm data in Figure 2 show that nearly $90\%$ of the 
sources follow a well defined sequence which is almost parallel to the reddening line.
However, 41/313 of the sources ($13\%$) have much bluer \it{(I-J)} \rm colours and lie to 
the left of the arbitrary line,  parallel to the reddening line, which is plotted in 
Figure 2. An 
initial suspicion that this might be due to poor photometry was disproved by observing that 
the same two sequences are followed in a subset of 92 sources located more than $\sim 2$ 
arcminutes away from the cluster core, where photometry is not compromised by bright 
nebulosity (not shown). Since we have selected filters with no strong low-excitation 
emission lines the obvious interpretation is that the 
blue colours are due to scattered light from those objects which are very young or are 
viewed 
close to the plane of the accretion disk. This interpretation is confirmed by comparison 
with the list of proplyds detected via emission line imaging with the \it{Hubble Space 
Telescope (HST)} \rm in O'Dell \& Wong (1996). Only 11/41 were detected in their relatively
shallow optical surveys, which suffer from greater extinction and do not overlap precisely 
with our survey. However, 10/11 are listed as proplyds, and 1/11 is listed as a star. The 
star is presumed to illuminate circumstellar matter which was not detected by \it{HST} \rm 
because it does not receive sufficient UV radiation from the central O-type stars, or 
because the proplyd ``tail'' lies too close to the line of sight. These 10 proplyds are 
marked with crosses in Figure 2.

	The 41 blue sources are widely distributed throughout the survey region and
do not display obviously unusual (J-H) colours, only a weak tendency to be bluer than the 
rest. The $I$ vs. \it{I-J} \rm diagram appears to be an efficient way of detecting proplyds 
at large distances from the photoionising O stars. However, the effect of circumstellar 
matter on infrared colours is not obvious, since it depends on the orientation of the 
system and distribution of matter in a complicated way (eg. Kenyon et al.1993). The 
observed (J-H) colours may well be different from the photospheric colours in these systems 
so we have excluded all the blue sources from our dereddening analysis. It is likely that 
some sources in the red group also have slightly modified colours due to anomalous 
extinction 
(i.e. both absorption and scattering) but this is not expected to be significant for most 
sources of age $\sim 1$~Myr, since the spectral energy distributions of T Tauri stars are 
usually well fitted by a Planckian (photospheric) function at wavelengths between the visible 
and the thermal infrared (eg. Rydgren et al. 1976, Wilking et al. 1989). The 202 faint sources 
detected only at $J$ and $H$ cannot be probed for 
anomalous colours in this way and probably include some sources which are therefore 
inaccurately dereddened. However, the proportion of anomalous sources will be lower than 
$13\%$ in this group because the blue sources are easier to detect at I band. Those 
sources in Figure 1(a) with m$_{J} > 16.2$ which are detected in all 3 filters are almost
exclusively members of the blue group. Hence, blue sources detected only at $J$ and $H$
will be very few except below the $I$ band detection limit which corresponds to 
$m_{J} \approx 18$. A likely example of one such faint source at lower left in Figure 1(a-b)
is Orion 131-047 (adopting the O'Dell \& Wen (1994) naming convention), for which m$_{J} = 18.22$ 
and \it{(J-H)} \rm= 0.39. This is an unrealistically blue colour for such a low luminosity source (which 
has accurately measured fluxes) so it is likely that the \it{(J-H)} \rm colour has been reduced 
by at least 0.2 magnitudes by scattering effects.

\subsection{Extinction Law}

	Many observers have found evidence for an anomalous extinction law in the Trapezium 
at optical and infrared wavelengths (eg. Davis et al.1986; Cardelli, Clayton and Mathis 
1989, hereafter CCM). We adopt the R$_{V}=5.5$ extinction law of CCM, calculated for 
$\theta_{1}$C Ori using data at similar wavelengths to those observed here. The infrared 
extinctions for this law are A(I$_{C}$)=0.643,  A(I$_{U}$)=0.583, 
A(J)=0.334, A(H)=0.214 for A(V)=1. The effect of the unusually high R$_{V}$ is to increase 
the slope of the reddening lines, resulting in higher derived luminosities and masses for 
sources with significant extinction. However, the change in near infrared reddening is quite 
small because, as showed by CCM, all extinction laws converge at wavelengths 
$\gtsimeq 0.9~\mu$m, such that A($\lambda$)/A(I) is similar in all known clusters. 
Only the optical extinction is greatly modified and we are not concerned with this. Hence, 
we use $A(J) = 2.783 E(J-H)$, which compares with $A(J)=2.364 E(J-H)$ for the R=3.05 
interstellar extinction law derived from Whittet (1990) and $A(J)=2.543 E(J-H)$ which 
follows from the oft-quoted $\lambda^{-1.8}$ law for near infrared interstellar extinction. 
This convergence of extinction laws in the near infrared is fortunate, since we doubt
whether a common extinction law will apply to all the stars in a given star
formation region, especially in Orion where the nebulosity has such complex spatial 
structure.

\subsection{Dereddening Procedure}

	For the $IJH$ detections, we have used the unusual procedure of dereddening to an 
empirically derived curve in the \it{(I-J) vs.(J-H) \rm two-colour diagram, rather than 
to a theoretical curve in a colour-magnitude diagram. This was for two reasons: firstly, 
theoretical models are at an early stage of development for such young substellar
sources and appropriate colour predictions have yet to be published at the time
of writing; secondly, theoretical colour predictions are subject to the significant
uncertainty in the absolute temperature-colour calibration which we referred to in 
Section 1.
The empirical curve (Figure 3(a)) was derived from a fit to observations of M and L dwarfs 
of near solar metallicity from Leggett et al.(1998) and UKIRT stellar standards of unknown 
metallicity for \it{(I-J)} \rm $<1$, where there is no known metallicity dependence. A 
possible
flaw in this approach is that young stars and brown dwarfs have  lower surface 
gravities than main sequence objects of the same effective temperature. However, the colour
predictions by Baraffe et al.(1998) for young low mass stars indicate that much larger 
changes
in $log(g)$ ($>1$ dex) have a negligible effect on these colours. The B97 models also 
indicate that $log(g)$ at 1~Myr does not approach red giant values for the masses considered 
here, so the empirical curve is not likely to be far in error. A cubic 
polynomial was used, with a Gaussian addition to fit the peak at \it{(I-J)} \rm $\approx 1.0$. 
The form of the relation is a good match to the plots of Bessell \& Brett (1988) and Leggett 
(1992). 

	The results for the 272 $IJH$ sources which are not anomalously blue in \it{(I-J)} \rm
are plotted in Figure 3(b). 95 sources have double valued solutions but only 1 solution 
has a plausible colour and flux in nearly every case. The handful of uncertain choices are low 
mass stars near the bump in the curve at $(I-J) \approx 1.0$, where the two solutions lie close 
together in \it{(J-H)} \rm and derived luminosity but differ substantially in \it{(I-J)} \rm and 
hence derived temperature. These ambiguous sources are listed as such in Table 1. We are 
confident that the dereddened \it{(J-H)} \rm colours are accurate to $\pm$ 0.1 mag in nearly 
every case, given the weak temperature dependence of this colour; a standard error of 0.05 mag is 
estimated, due to measurement error and uncertainties in the empirical curve and dereddening law. 
The $J$ band extinction correction should therefore have typical uncertainties of $\pm 0.14$ mag, 
which is small enough to produce a useful Luminosity Function. The dereddened \it{(I-J)} \rm 
colours are more sensitive to any errors in the process, such that the standard error is
approximately 0.25 mag. However, the very strong temperature dependence of the \it{(I-J)} \rm 
colour means that this leads to only a modest uncertainty in derived values of $T_{eff}$ 
(q.v Section 4.3).

	The 202 $JH$-only detections were dereddened to the theoretical track shown in
the colour-magnitude diagrams (Figure 1(a-b), which is a simple linear fit to an $L$-$T_{eff}$ 
relation (taking the average of the DM98 and B97 predictions at 1~Myr), and a 
$T_{eff}$-\it{(J-H)} \rm relation (taking the average of the Wilking et al (1999) and Baraffe et 
al.(1998) relations for main sequence stars, which agree to 0.02 mag). Only the B97 and
Baraffe predictions extend to the faintest magnitudes and lowest temperatures (\it{(I-J)}$>3.3$) 
\rm. In this region the \it{(J-H)} \rm colour changes more rapidly with T$_{eff}$ and the 
uncertainties increase.

\section{Interpretation}

\subsection{Luminosity Function and IMF}
	
	The Luminosity Function (LF) is plotted for all sources detected at $H$ band with 
m$_{H} > 
12.25$ in Figure 4(a). The function declines from a strong peak at m$_{H} \le 12.5$, which is not 
well measured here due to saturation. Zinnecker, McCaughrean \& Wilking (1993) and Ali \& Depoy 
(1995) observed the equivalent K band function 
and found a peak at m$_{K}$=11-12, the function declining to fainter magnitudes but flattening off
and possibly rising beyond m$_{K}$=14. In our data a strong peak at $H$=16.5 is apparent, which
probably has physical significance given that the function is based on more than 500 sources,
and any real features are blurred by extinction of typically 1 magnitude at H band.
The peak exists independently of the magnitude binning and is seen in Figure 1(b) as a clump of
sources with 16$<m_{H}<$17. A corresponding feature exists in the J band LF at J=17.5.  
The completeness falls gradually, due to the variable nebular surface brightness but is
estimated at $> 90\%$ to m$_{H}$=18.0, as evidenced by the small secondary peak there.

The observed function is converted to the absolute Luminosity Function shown in Figure 
4(b), including only the dereddened sources (i.e. excluding H band only detections and blue
$IJH$ sources. $M_{bol}$ is determined for each source using $J$ band magnitudes and 
bolometric corrections (these being well established at $J$) and a distance modulus of 8.22. 
The bias due to necessarily retaining some faint, anomalously blue sources due to lack of 
$I$ band data is only significant below m$_{J}$=18, which is approximately the completeness limit 
(the nebulosity reduces sensitivity more in the $I$ and $J$ bands than at $H$.) 
We adopted the following bolometric corrections, derived from a simple fit to  
the relations between T$_{eff}$, \it{J-H} \rm and BC$_{J}$ of Wilking et al.(1999) (and Baraffe 
et al.(1998) for T$_{eff} > 3500~K$) and using the DM98 1~Myr isochrone to connect T$_{eff}$ to
Luminosity:  

\vspace{-4mm}
\begin{eqnarray}
	BC_{J} = 1.955;    J_{dr} > 13.77 \\
	BC_{J} = 0.1583 J_{dr} -0.2248;  J_{dr} < 13.77
\end{eqnarray}

where $J_{dr}$ is the dereddened $J$ magnitude and the BC$_{J}$ refers to the UFTI filter,
which has smaller bolometric corrections than the CIT filter.

Figure 4(b) shows a primary peak at M$_{bol}=6$ and a small secondary peak at M$_{bol}=10.5$.
Incompleteness is significant for M$_{bol}<6.2$, due to saturation of bright sources, and
for M$_{bol}>11.75$, which corresponds to $J > 18$. The secondary peak corresponds roughly to
the peak at $H=16.5$ in Figure 4(a). We convert the LF to the IMF using the tracks of B97
and DM98. To remove bias due to non-detection of highly reddened sources
we include only sources with $(J-H)_{obs} < 1.5$. The results in Figure 5 therefore 
represent an unbiased sample of the IMF complete to $log(M/M_{\odot})=-1.5$. In a log-log 
plot, both IMF's show  a fairly flat stellar function which has a tendency to fall slightly
into the brown dwarf regime. Both IMFs also show some indication of a rise beyond the 
completeness limit but deeper observations will be needed to quantify this.

	The discovery of a large population of brown dwarf candidates contrasts with previous 
surveys (eg. Hillenbrand \& Hartmann 1998) which have concluded that few substellar objects exist.
This conclusion appears to have been based on the well-established decline in the LF
beyond the principal peak at m$_{K} \approx 11.5$. This survey is the first to go deep
enough at infrared wavelengths to detect the secondary peak in the LF. 

\subsection{A Cut-off in the Luminosity Function ?}

	The absence of faint blue sources (see Section 3.1) is well established by
the inclusion of the $H$ band upper limits. This may indicate a sharp turn-down in the
LF, and perhaps a cut-off, at a level corresponding to about 8 Jupiter masses. However,
such sources would be close to the survey sensitivity limit (which we believe to lie
just below 5~M$_{Jup}$) particularly if their intrinsic \it{(J-H)} \rm colours are redder than 
we expect. Moreover, the B97 mass-luminosity relation becomes steeper below about 8 M$_{Jup}$, 
so a turn down in the LF would occur even for a flat IMF. Hence a deeper survey will be needed 
to confirm the reality of the fall in the IMF. The least massive detection with good photometry 
is Orion 023-115, which has J=19.38, \it{(J-H)} \rm=0.97 and a derived
mass of $8.4^{+1.4}_{-2.7}$ M$_{Jup}$ ($8.0^{+1.3}_{-2.6} \times 10^{-3} M_{\odot}$) for an 
age of 1~Myr, using the B97 tracks. The quoted $+/-$ uncertainties refer to alternative ages 
of 2~Myr and 0.3~Myr respectively. None of the faint sources in Figure 1(a) with highly 
uncertain $J$ band fluxes have a lower derived mass. If real, the turn-down 
might be attributed to a minimum Jeans mass for gravitational cloud core collapse, below 
which star formation cannot occur. Alternatively, it may be due to the ending of star 
formation in the cluster before extremely low mass cloud cores had time to collapse, since 
this process is believed to take longer in less massive cores. This may be attributable to 
dispersal of dense molecular gas by the photoionising O-type stars at the centre of the 
cluster. The IAC group apparently find no sign of the turn down in the LF at planetary 
masses in the neighboring $\sigma$ Orionis cluster, so we favour the second explanation. If 
correct, this explanation may still be significant with regard to any galactic population of 
free-floating planets: most star formation is believed to occur in high mass star 
formation regions like the Trapezium or M16, with the consequence that free-floating planets 
may be relatively rare. However, the effectiveness of this mechanism for reducing planet 
formation efficiency will vary depending on local conditions.

\subsection {Effective Temperatures}

	The dereddened \it{I-J} \rm colours are converted to the effective temperatures in
Table 1 using the models of Baraffe et al.(1998). The derived values of T$_{eff}$ are in 
reasonable agreement with the predictions of B97 and DM98, in which $2600 < T_{eff} < 2900~K$ for 
objects of brown dwarf mass at 1~Myr. However a few sources in this mass range are detected
with dereddened $(I-J) \ge 3$, which implies $T_{eff} \le 2500$~K, at least for main sequence 
objects. We estimate that our derived temperatures are accurate to $\pm 200~K$, leaving
aside the uncertainty in the absolute T$_{eff}$ vs. \it{I-J} \rm relation and assuming this
is not altered significantly by the youth of the sources. In any case the \it{I-J} \rm colours
should provide a useful guide to relative temperatures.

\subsection {Potential Problems}

	We have carefully avoided several potential problems in studies of young clusters, 
such as cluster membership, infrared excess and line emission but some serious issues
remain. We consider each in turn. (1) Foreground contamination is minimal in the 
tiny area surveyed, introducing perhaps 5 red dwarfs into the sample over a range of about 3 
magnitudes, using the stellar space densities of Tinney (1993). (2) Background contamination is 
removed by the dark backdrop of OMC-1 (see Hillenbrand \& Hartmann 1998). Our avoidance of the K 
band filter removes any chance of seeing through OMC-1 at the faint limits. (3) Infrared excess 
due to hot dust is not believed to be significant at H band, since the spectra of T Tauri stars 
are well fitted by black bodies at this wavelength. (4) Line emission was minimised by the
choice of filters. (5) Variation in the infrared extinction law will probably be small
(see Section 3.4) and the effect on the derived IMFs is minimised by excluding highly
reddened sources. (6) Scattering is a potentially serious problem. Even sources without
anomalous colours in the $I$ vs.\it{(I-J)} \rm diagram may have distorted \it{(J-H)} \rm colours 
and fluxes. As noted in Section 3 however, photospheric flux dominates the spectral energy 
distributions of most T Tauri stars for 0.6~$\mu$m $ < \lambda < $2~$\mu$m) so significant 
distortion of broad band colours and fluxes is unlikely to be common. This should be investigated 
in future by searching for the polarisation signature of scattered light. (7) The evolutionary 
tracks for substellar objects are in an early stage of development, which leads to a significant 
uncertainty in derived masses. However the fairly close similarity of the B97 and DM 98 tracks, 
for luminosity and $T_{eff}$, is encouraging.

\section {Conclusions}

	A large population of brown dwarf candidates is detected in the Trapezium Cluster
and a small population of objects with planetary masses. We have confidence that these are 
true cluster members and, though many uncertainties exist in deriving the masses, they are not 
likely to be large enough to cause misclassification of low mass stars as low mass brown dwarfs 
or free-floating planets. The derived IMF is fairly flat on a log-log plot at low stellar masses 
but declines slightly at brown dwarf masses, indicating that brown dwarfs are a little less
numerous than stars. There is a possible small peak near $\sim 0.02$M$_{\odot}$, which is below 
the completeness limit. Approximately 13 planetary mass objects are detected but none with
M$< 8 \times 10^{-3}$M$_{\odot}$. We suggest that this is due to dispersal of the star-forming 
cloud by the photoionising O-stars before such objects had time to form. 

	Approximately 13\% of sources have anomalously blue \it{(I-J)} \rm colours, which we
attribute to scattering from circumstellar material. These blue excesses are strongly
correlated with detection as 'proplyds' by \it{HST} \rm, so this colour selection may prove
to be a powerful new tool for detecting sources with circumstellar envelopes. The
effect of scattering on the colours of the general population should be investigated 
via polarimetry and spectroscopy.

\vspace{4mm}
\large{\textbf{Acknowledgements}}\textmd{}\normalsize

        We wish to thank the staff of UKIRT, which is operated by the Joint 
Astronomy Centre on behalf of the UK Particle Physics and Astronomy Research 
Council (PPARC). Particular thanks are due to Sandy Leggett and Andy Adamson
for providing information on the colours of cool stars and for carrying out
reactive observations for us. We also thank the Panel for the Allocation of Telescope 
Time for providing us with reactively rescheduled observing time. 
We thank Juliette White for helping us with the dereddening procedure
and we are grateful to the referee, Richard Jameson, for useful and timely comments.  
PWL is grateful for support by PPARC via a Post Doctoral Fellowship at the 
University of Hertfordshire.

\vspace{1cm}

\large{\textbf{References}}\textmd{}\normalsize
\setlength {\parskip} {2mm}
\setlength {\parindent} {0mm}

Ali B., \& Depoy D.L. 1995, AJ, 109,709 

Baraffe I., Chabrier G., Allard F., \& Hauschildt P.H. 1998, A\&A, 337,403 

Bessell M.S., \& Brett J.M., PASP, 100,1134

Burrows A. 1997, ApJ, 491,856

Cardelli J.A., Clayton G.C., \& Mathis J.S. 1989, ApJ, 345,245

Comeron F., Rieke G.H., Burrows A., Rieke M.J. 1993, ApJ 416,185

Comeron F., Rieke G.H., \& Rieke M.J., ApJ 473,294

D'Antona F., \& Mazzitelli 1997, MmSAI, 68,607 

Davis D.S., Larson H.P., \& Hofmann R., 1986, ApJ, 304,481

Hillenbrand L.A. 1997, AJ 113,1733 

Hillenbrand L.A., \& Hartmann L.W. 1998, ApJ, 492,540

Kenyon S.J., Whitney B.A., Gomez M., \& Hartmann, L., 1993, ApJ 414,773 

Leggett S.K., Allard F., \& Hauschildt P.H. 1998, ApJ 509,836

Luhman K.L., \& Rieke G.H., 1998, ApJ 497,354

Luhman K.L., Rieke G.H., Lada C.J., \& Lada E.J., 1998, ApJ 508,347

Nakajima T., Oppenheimer B.R., Kulkarni S.R., Golimowski D.A., Mathews K.,
\& Durrance S.T., 1995, Nature 378,463

O'Dell C.R., \& Wong K. 1996, AJ 111,846 

Rebolo R, Zapatero-Osorio M.R., \& Martin E.L., 1995, Nature 377,129

Roche P.F., \& Lucas P.W., 1998, on-line, www-astro.physics.ox.ac.uk/~pwl/camera.html

Rydgren A.E., Strom S.E., \& Strom K.M., ApJS 30,307

Tinney C.G. 1993, ApJ, 414, 279

Whittet D.C.B., 1992, Dust in the Galactic Environment. Institute of Physics, Bristol.

Wilking B.A., Greene T.P., \& Meyer M.R. 1999, AJ 117,469 

Wilking B.A., Lada C.J., \& Young E.T. 1989, ApJ 340,823

Williams D.M., Comeron F., Rieke G.H., \& Rieke M.J., 1995, ApJ 454,144

Zinnecker H., McCaughrean M.J., \& Wilking B.A. 1993, in ``Protostars and Planets III'',
p429, edited by E.H.Levy \& J.I.Lunine, pub. Tucson: University of Arizona Press.



\pagebreak

Figure 1: (a) J vs. (J-H) plot. Open circles are highly uncertain data points. 
The dotted line is an approximate zero reddening track (see text). The solid lines are
parallel to the A(V)=7 reddening vector and divide the population into stars, brown 
dwarfs and 
planets, using the B97 prediction and an age of 1~Myr. The dashed lines correspond to
the 0.3~Myr and 2~Myr predictions, indicating the effect of the age spread on the
classification. The effect is similar at the planetary boundary.\\ 
(b) H vs (J-H) plot. This includes upper limits for sources with $J>20$, which
confirms the paucity of faint blue sources.

Figure 2: I vs (I-J) plot. Anomalously blue sources lie to the left of the arbitrary
line parallel to the reddening track. Of the 11 blue sources detected by HST, 10 
are proplyds, plotted as crosses.

Figure 3: (a) Empirical 2-colour curve fitted to the plotted observations. Two example
dereddening tracks are also shown, indicating the possibility of double-valued 
solutions.\\ (b) results of dereddening Orion data. 

Figure 4: (a) Observed H band luminosity Function. The equivalent
J band function is overplotted as a dashed line. Both functions are complete to
approximately magnitude 18. 
\\ (b) Dereddened Luminosity Function, complete between magnitudes 6.2 and 11.75.

Figure 5: IMFs for Burrows 1997 and DM98 tracks, complete to log(M/M$_{\odot})=-1.5$.
Errorbars are plotted assuming Poisson statistics. The IMF appears to fall slowly 
into the Brown Dwarf regime, but rises again below the completeness limit. 

\pagebreak
\pagestyle{empty}
\begin{figure*}[thbp]
\begin{center}
\begin{picture}(200,400)

\put(0,0){\includegraphics{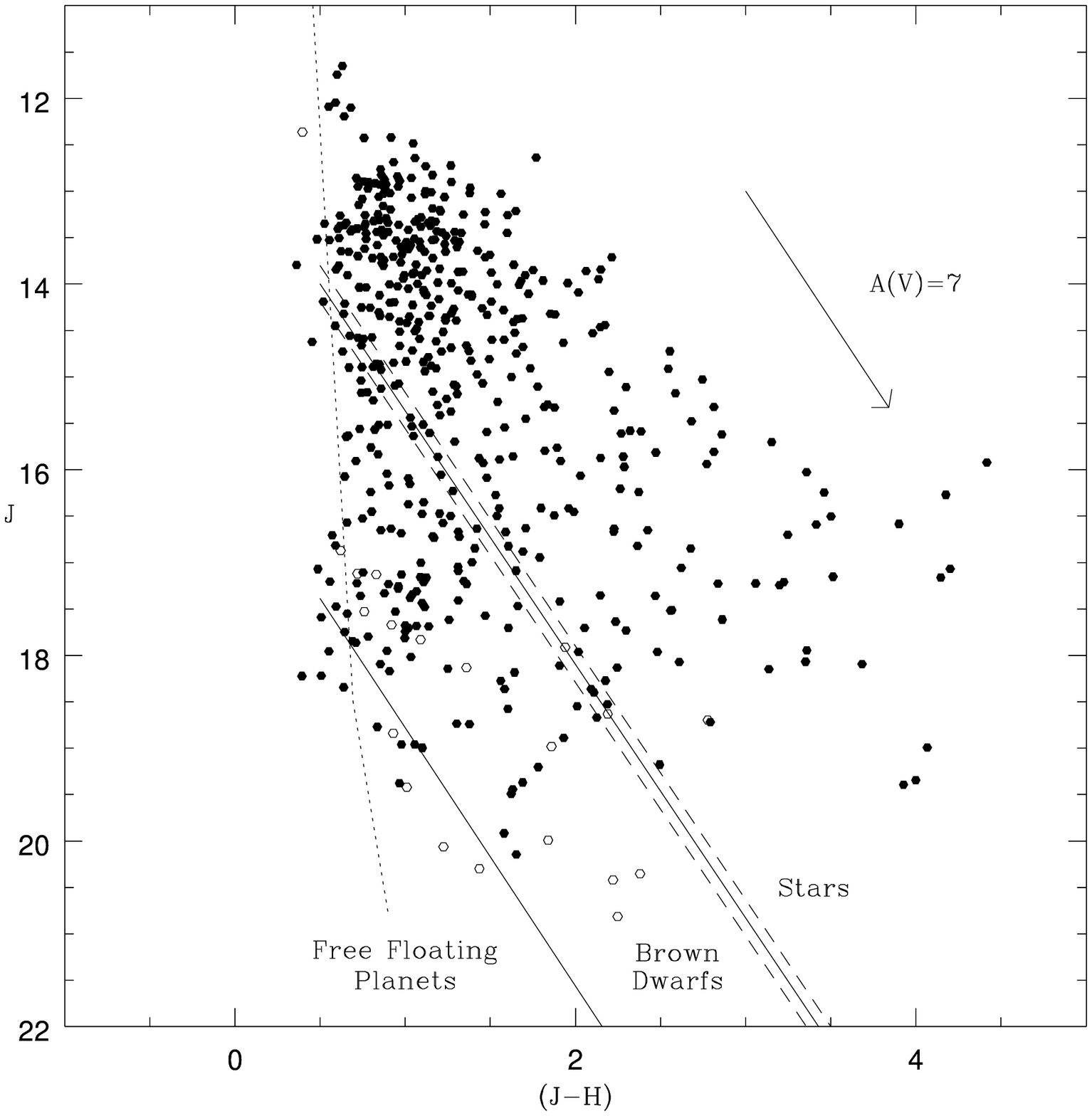}}

\put(0,0){\includegraphics{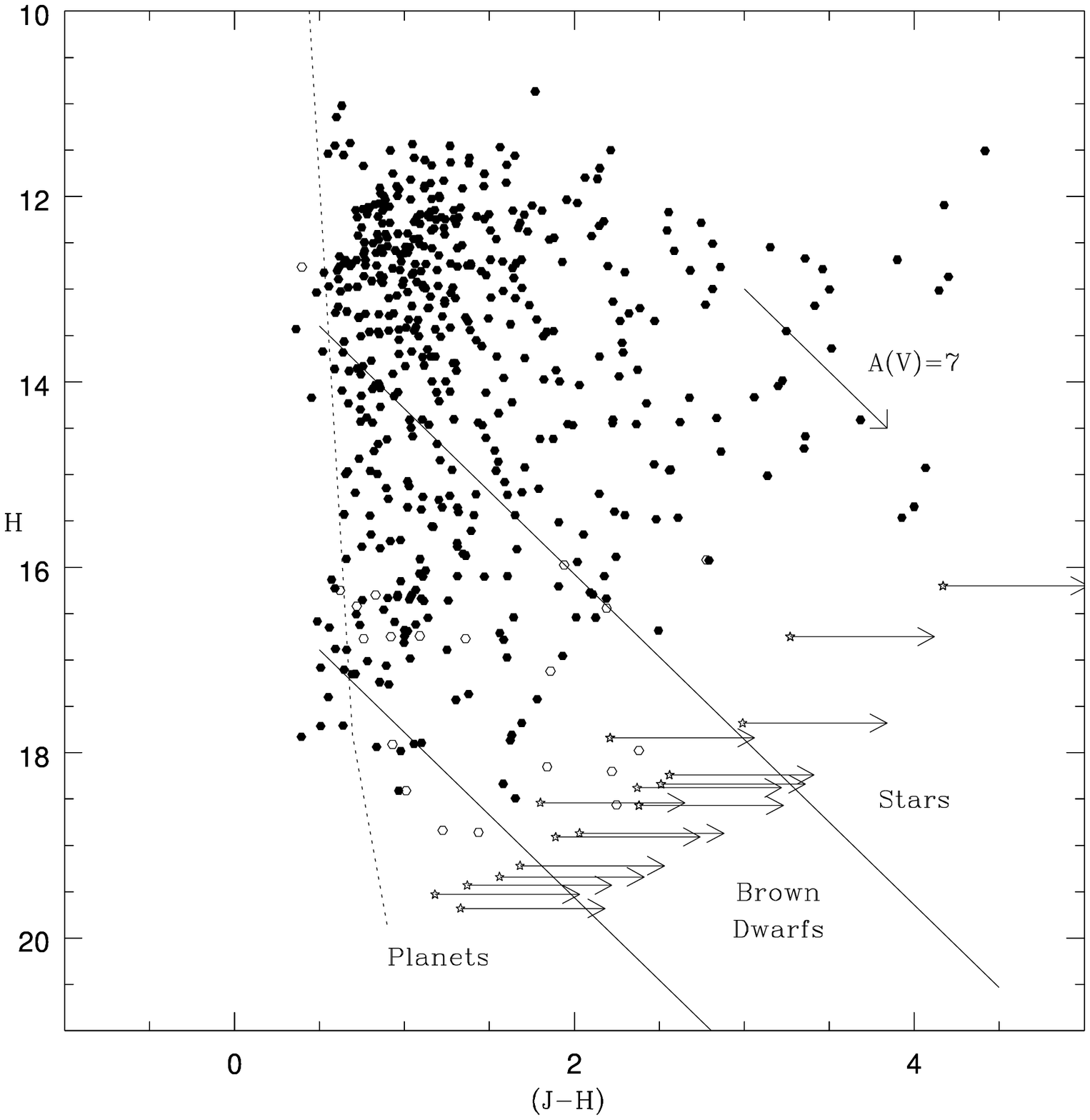}} 

\end{picture} 
\end{center}
\end{figure*}

\pagebreak
\begin{figure*}[thbp]
\begin{center}
\begin{picture}(200,400)

\put(0,0){\includegraphics{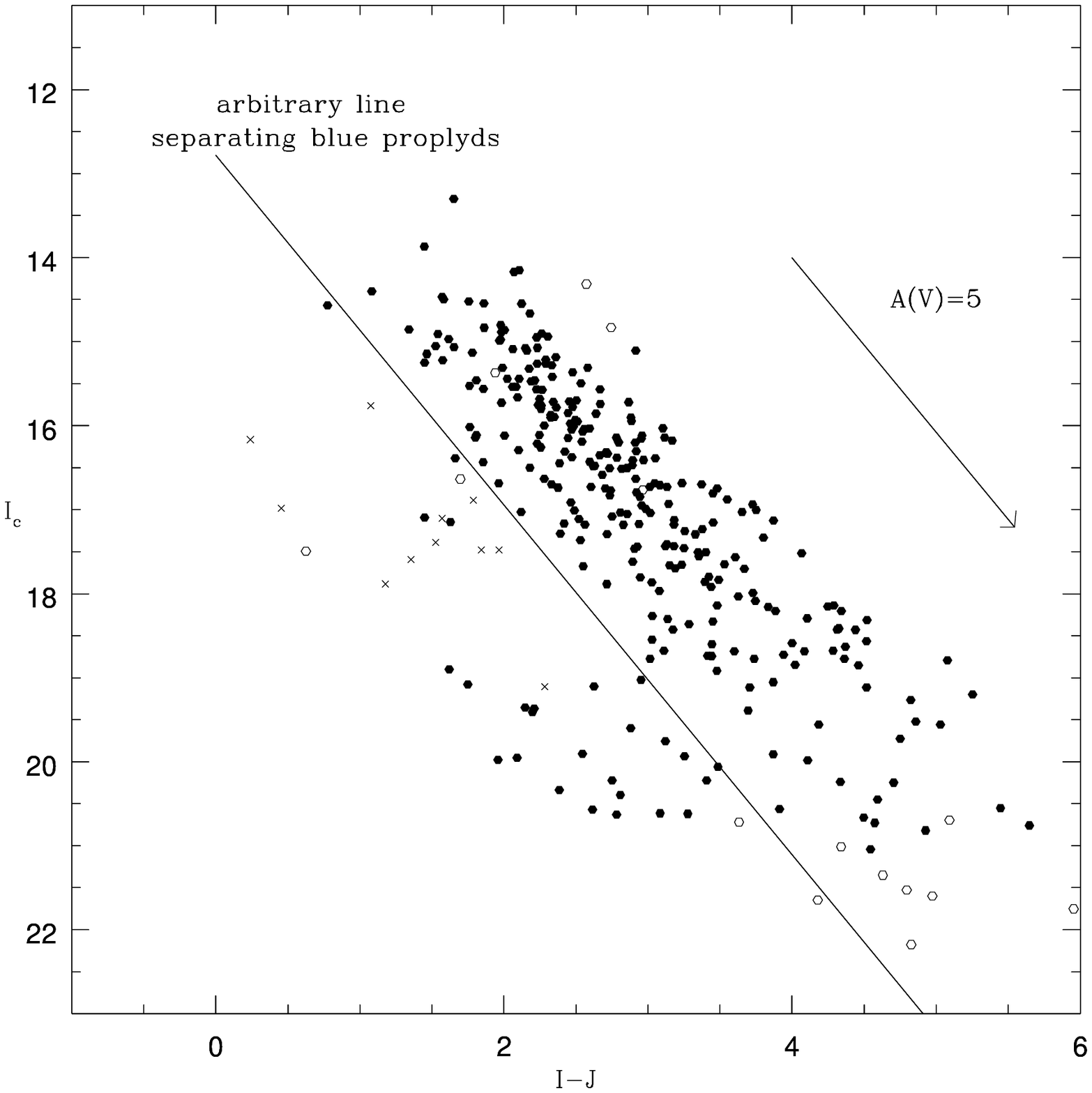}}

\end{picture} 
\end{center}
\end{figure*}

\pagebreak

\begin{figure*}[thbp]
\begin{center}
\begin{picture}(200,200)

\put(0,0){\includegraphics{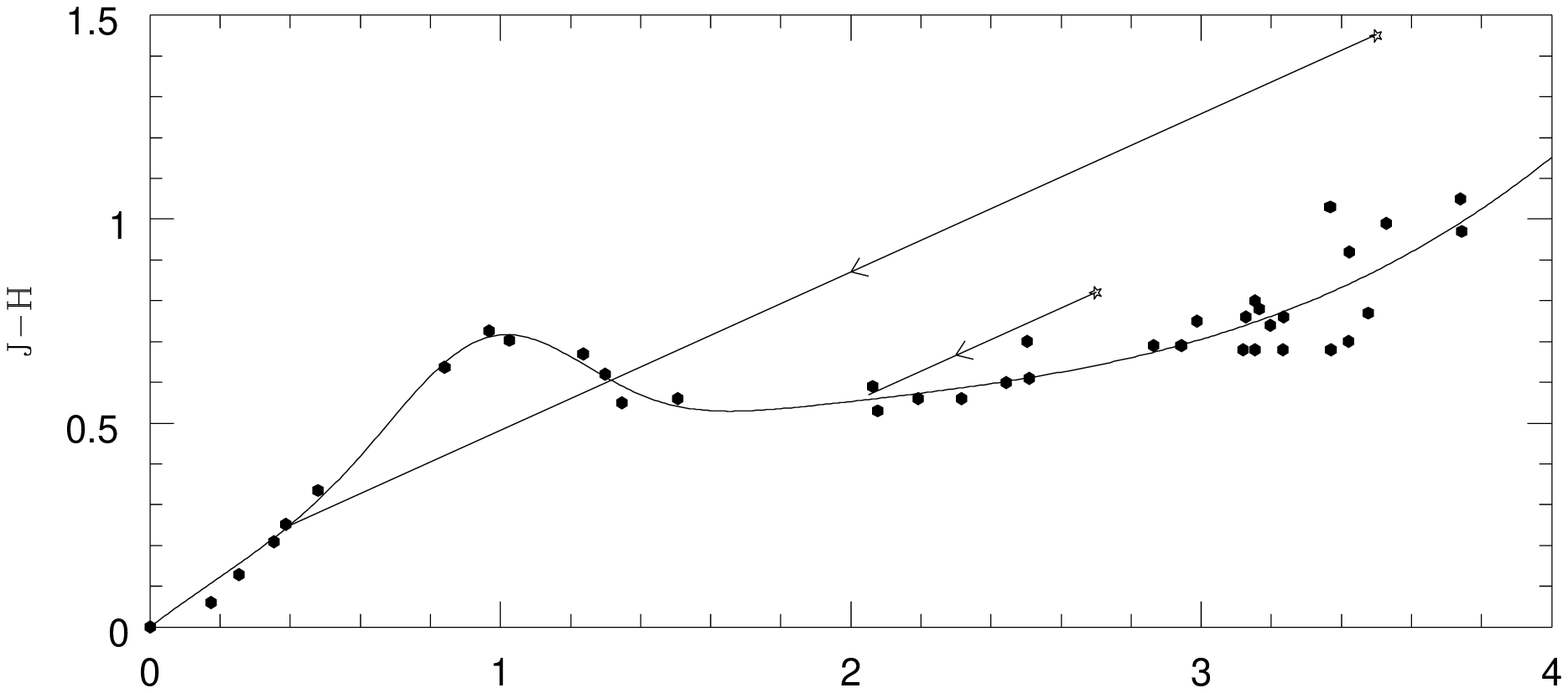}}

\put(0,0){\includegraphics{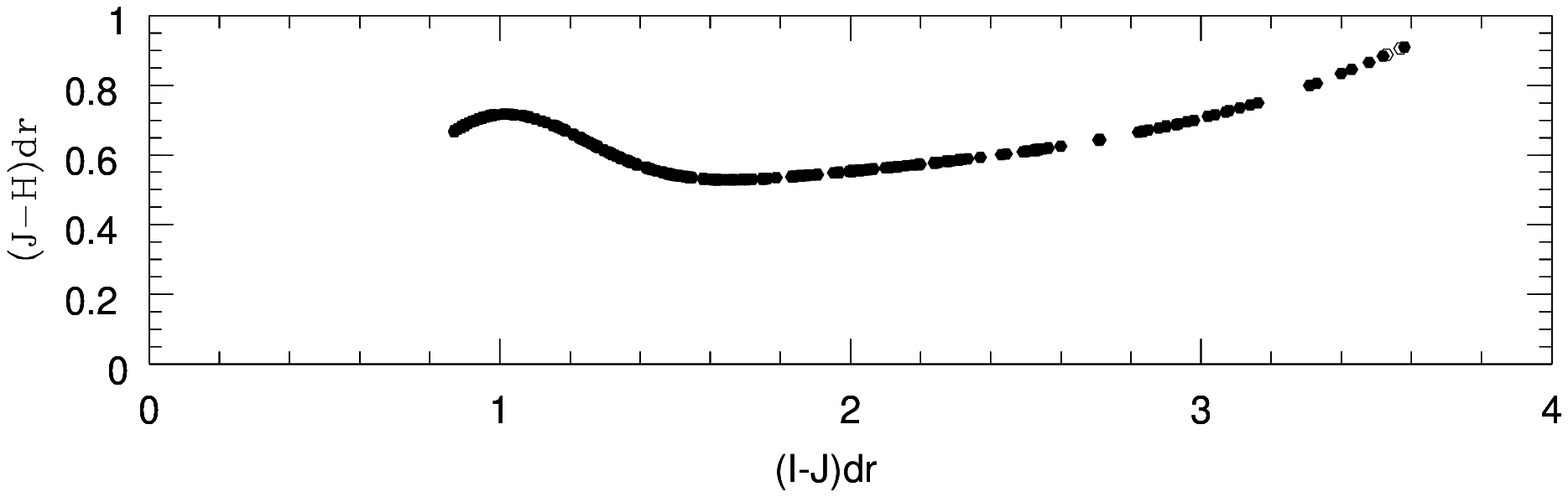}} 

\end{picture} 
\end{center}
\end{figure*}

\pagebreak
\begin{figure*}[thbp]
\begin{center}
\begin{picture}(200,400)

\put(0,0){\includegraphics{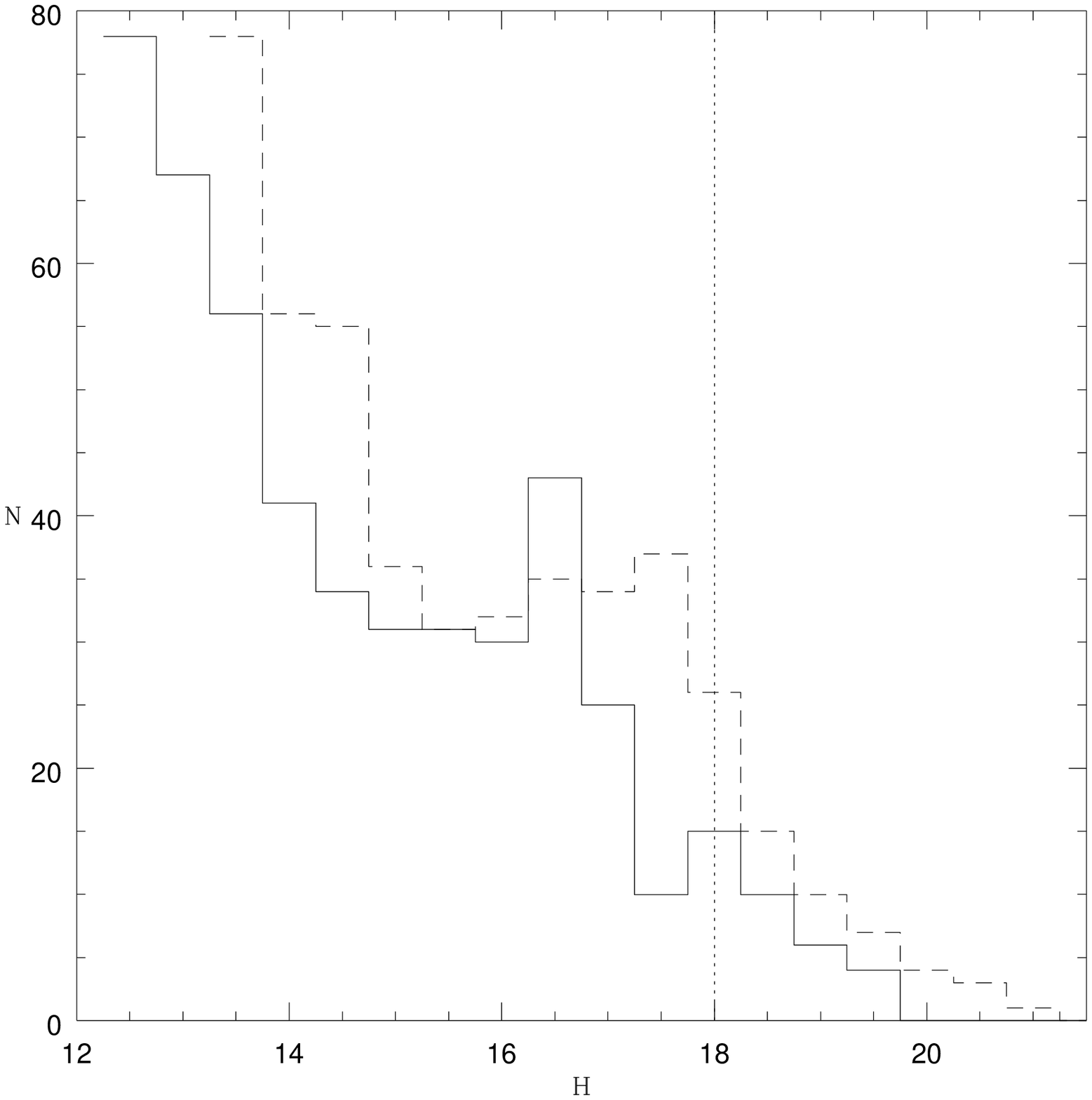}}

\put(0,0){\includegraphics{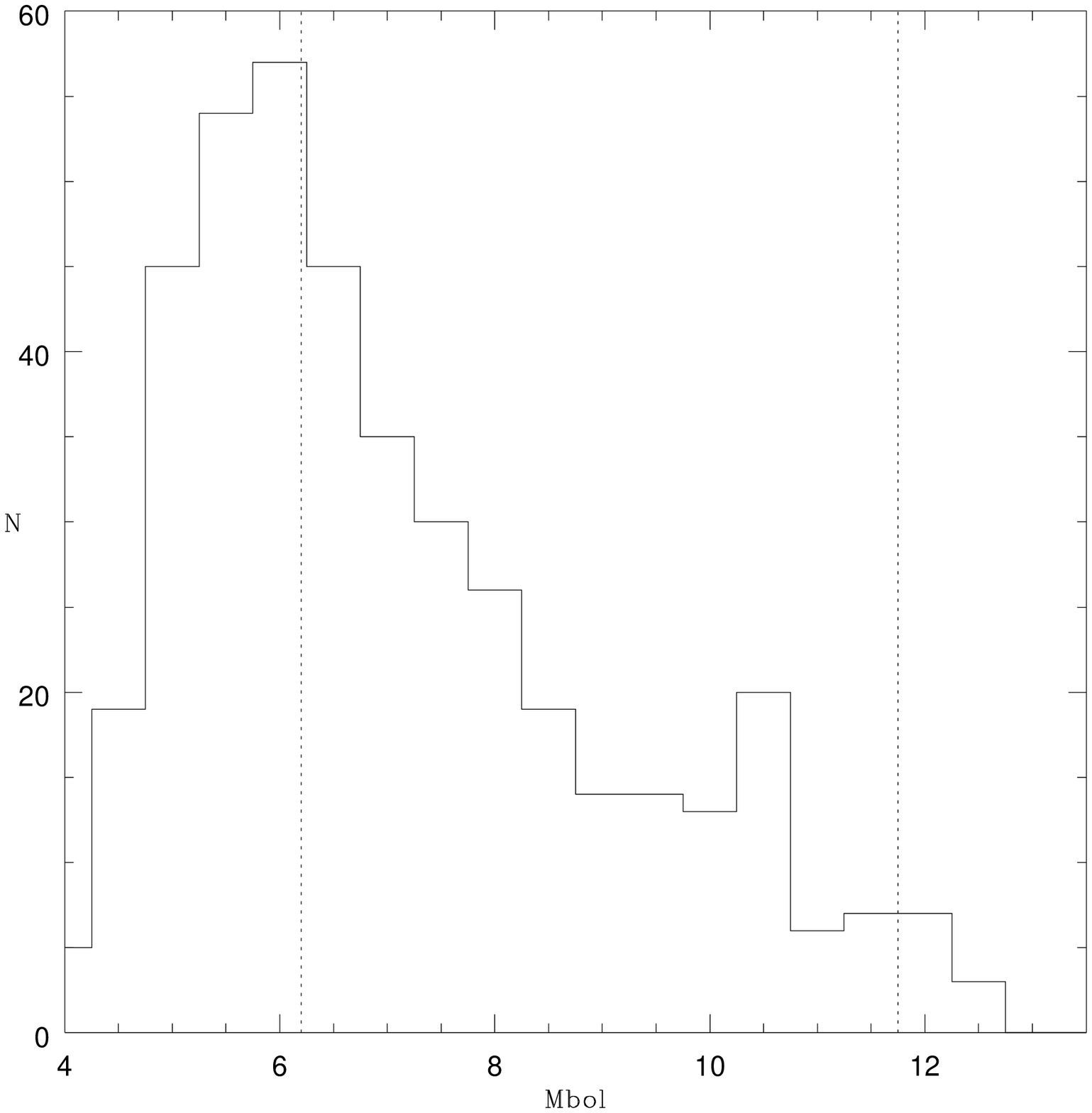}} 

\end{picture} 
\end{center}
\end{figure*}

\pagebreak
\begin{figure*}[thbp]
\begin{center}
\begin{picture}(200,400)

\put(0,0){\includegraphics{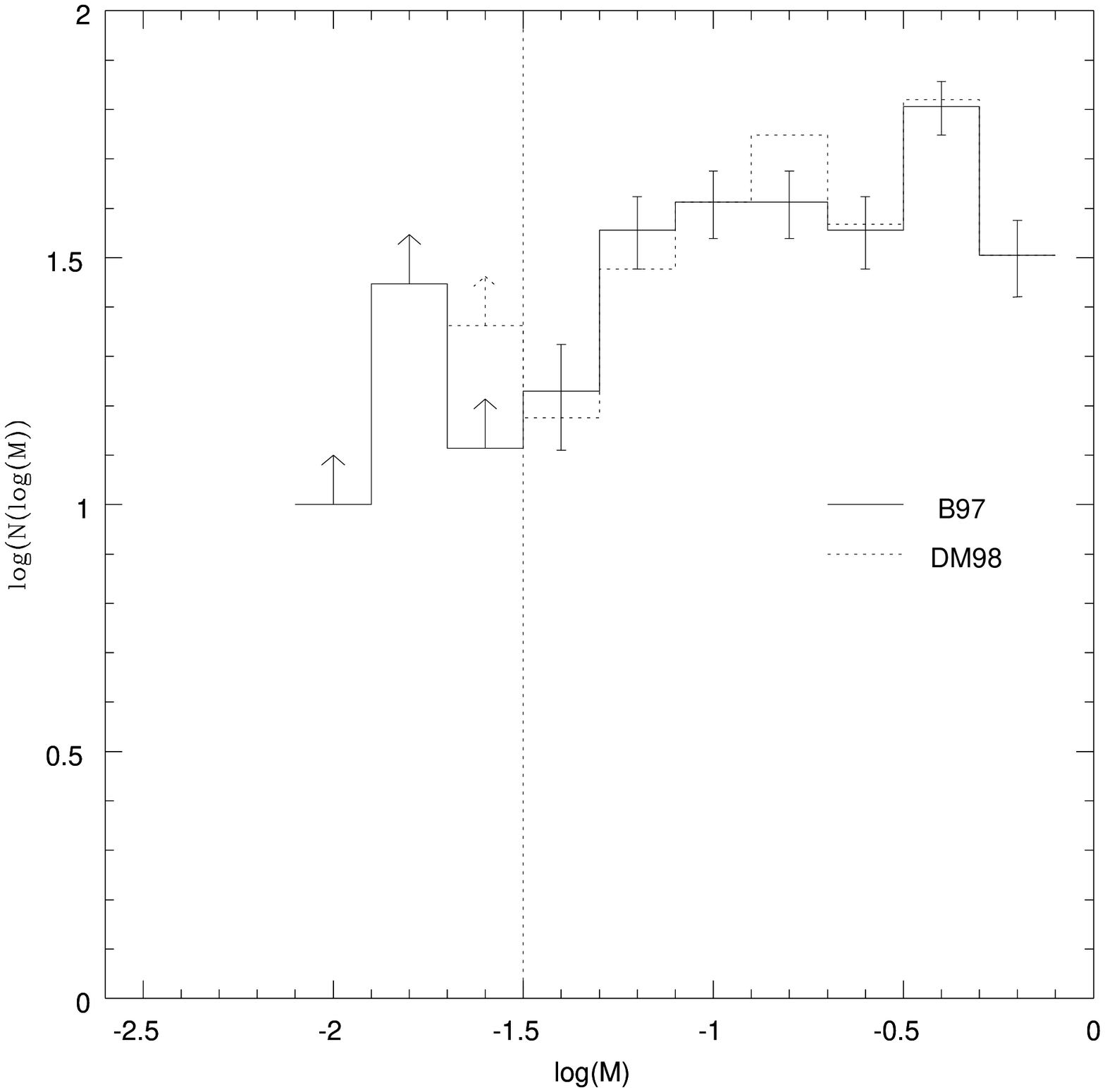}}

\end{picture} 
\end{center}
\end{figure*}

\end{document}